\documentclass{eptcs}

\usepackage{upgreek,latexsym, amssymb, amsmath, amsfonts, mathrsfs} 

\usepackage{ndj}

\newcommand{\mv}[2]{{\tt move}_{#1,#2}}

\newcommand{\spec}{s^1_1}

\title{A Swiss Pocket Knife for Computability
}
\author {Neil D. Jones
\institute{Computer Science Department\\
University of Copenhagen\\
2100 Copenhagen, Denmark}
\email{\quad neil@diku.dk}}
\def\titlerunning{Swiss pocket knife}
\def\authorrunning{N.D. Jones}

\begin{document}


\newpage

\maketitle

\begin{abstract}

This research is about operational- and complexity-oriented aspects of  classical foundations of  computability theory.
The approach is to re-examine some  classical theorems and constructions,  but with new criteria for success that are natural  from a  programming language perspective.

Three cornerstones of computability theory are the  {\em S-m-n} theorem; Turing's  ``universal machine''; and  Kleene's second recursion theorem. 
In today's programming language parlance these are respectively partial evaluation, self-interpretation, and reflection.
In retrospect it is fascinating that  Kleene's 1938 proof is constructive; and in essence builds a self-reproducing program.

Computability theory originated in the 1930s, long before the invention of computers and programs. Its emphasis was on delimiting the boundaries of computability.
Some milestones include 1936 (Turing), 1938 (Kleene), 1967 (isomorphism of programming languages), 1985 (partial evaluation), 1989 (theory implementation), 1993 (efficient self-interpretation) and 2006 (term register machines).



The ``Swiss pocket knife''  of the title is a programming language  that allows efficient computer implementation   of  all three computability cornerstones, emphasising the third:  Kleene's second recursion theorem.
We describe experiments with a tree-based computational model aiming for both fast program generation and  fast execution of the generated programs. 

\end{abstract}

\section{Introduction}

\subsection{Context}

 {\em ``The grand confluence'' of the 1930s} (term due to Gandy \cite{gandy}) was a first major accomplishment of  computability theory: the
realisation that the classes of computable problems defined using Turing machines, lambda calculus, register machines, recursion schemes, rewrite systems,\ldots,  {\em are all identical}.
This led to a deep mathematical theory (recursive function theory) about the boundary between computable and uncomputable problems, e.g., by Kleene, Turing, McCarthy, Rogers \cite{kleenebook,kleene,mccarthy,minsky,rogers,turing}.
Three cornerstones of computability theory were identified: the  {\em S-m-n} theorem; the  ``universal machine'' (as  Turing called it); and  Kleene's second recursion theorem. 

What interests computer scientists, though, and what recursive function theory does {\bf not} account for,  is
the {\em time it takes to compute}  a function:
 the  {\em size and efficiency} of the programs involved.

\subsection{Contribution of this paper}  

Our research program is to re-examine classical computability theory constructions from an efficiency viewpoint.
Some accomplishments so far:   {\em partial evaluation} has applied one computability cornerstone, the {\em S-m-n theorem}, to program optimisation. The field of partial evaluation field is now  substantial, e.g., as documented by Jones, Gomard, Sestoft in \cite{JGS}.
This paper  focuses on another computability cornerstone: {\em Kleene's second recursion theorem} 
\cite{kleene}.

To study complexity aspects of the  classical recursion-theoretic results one needs good models of computation that take into account programs' running times, as well as their expressivity.
A stimulus for the current work was the elegant construction by Moss  in \cite{moss}, using the  novel {\tt 1\#} language  to prove Kleene's second recursion theorem. 

This paper  describes computer describes constructions and experiments with a tree-based computational model aiming for both fast program generation and  fast execution of the generated programs.
A programming language perspective has led to a better problem understanding, and improved asymptotic efficiency.
The net effect is that all three computability cornerstones can be efficiently implemented. 

Advances in the first two  made by a Copenhagen group and others include {\em constant-overhead self-interpretation}, documented  in \cite{JGS,stoc} and other places. 
Efficiency improvements  over  \cite{botik,moss,stoc} 
include a {\em constant-time implementation} of Kleene's second recursion theorem as by Bonfante and Greenbaum in \cite{bonfantegreenbaum}, and 
constant-time versions of the  constructions by Moss in \cite{moss}.

\section{Fundamentals}

\subsection{Recursive function theory}

{\bf Rogers' axioms}: an ``acceptable''  programming language consists 
of\footnote{We follow the line of  
Rogers' definition of an {\em acceptable enumeration of the partial recursive functions} \cite{rogers,machteyyoung}. Our variant: we write $\lsem p\rsem$ instead of $\phi_p$; and we use  programs and data from the two sets $\mathit{Pgm}, D$ instead of the natural numbers $\nats$.
}

\bi

\item Two sets, ${\it Pgms}$ and $D$ (of programs $p,q,e,\ldots$ and data $d,s,x,y,\ldots$), with ${\it Pgms} \subseteq D$.

\item A {\em semantic function}
$$
\lsem p \rsem^n : D^n \rightharpoonup D
$$
associating with each program and each $n$ a partial $n$-ary function. (We omit $n$ in $\lsem p \rsem^n$ when it is obvious from context.) The semantic function must have these properties:

\be
\item {\em Turing completeness}: a (mathematical partial) function $f: D^n \rightharpoonup D$ is computable {\em if and only if} $f=\lsem p \rsem^n$ for some program $p$.

\item {\em {\it Universal program} property}\footnote{
In  equations involving program semantics $=$ stands for equality of partial values, so $e_1=e_2$ means that either $e_1$ and $e_2$ evaluate to the same value, or both are undefined (in practice, meaning: nonterminating).}:
$$
\exists univ   \in \mathit{Pgms}\ \forall p   \in \mathit{Pgms}\ \forall d \in D \ (\ 
 \lsem p \rsem (d) =  \lsem univ \rsem (p,d) \ )
$$
\item{\em  {\it S-m-n} property (here for $m=n=1$)}:
$$
\exists \spec   \in \mathit{Pgms}\ \forall p   \in \mathit{Pgms}\ \forall s,d  \in D\ (\ 
 \lsem p \rsem (s,d) = \lsem \lsem \spec \rsem (p,s) \rsem (d) \ )
$$

\ee

\ei

\fl Restated in computer science terms:
a universal machine $univ$ is a {\em self-interpreter}; and 
an  {\it S-1-1} program $\spec$ is  a {\em partial evaluator} or
{\em program specialiser}.
(Remark:
Rogers has proven in  \cite{rogers} the remarkable result  that any two acceptable enumerations are computably  isomorphic.)

\subsection{Kleene's second recursion theorem}

Kleene's second recursion theorem (SRT for short) is an early and very general consequence of the Rogers axioms for computability. It clearly has a flavor of {\em self-application}, as it in effect asserts the existence of programs that can refer to their own texts.
The statement and proof are short,  though the theorem's implications are many.

\vair

\fl{\bf Theorem (Kleene, \cite{kleene})} For any acceptable  programming language,
$$
\forall p   \in \mathit{Pgms} \ \exists  p^*  \in \mathit{Pgms}\  \forall  d  \in D \ .\  \lsem p\rsem(p^*,d) = \lsem p^* \rsem (d).
$$

\fl In effect  this is a {\em program transformation}: it takes a 2-argument program $p$, and constructs from it a 1-argument program $p^*$. Program $p^*$ is sometimes called a ``fixpoint'', though it is not necessarily minimal, nor unique for the given $p$.\footnote{Rogers \cite{rogers} has an alternative version: that any computable total function $f$ from programs to programs has what could be called a ``syntactic fixpoint'': a program $p$ such that $\lsem p \rsem = \lsem f(p) \rsem$. It appears that Kleene's original version is in some sense more general than Rogers', cf. \cite{case}, so we only consider  the Kleene version.}

The theorem can be interpreted operationally, but was proven long before the first computers were built.
Applications of Kleene's theorem are many, and include:

\be

\item \label{app-self-reproducer}
 A {\em self-reproducing } program: Suppose  $\lsem p\rsem(q,d) = q$ for any $q$. Then for any input $d$,
$$\lsem p^*\rsem(d) = \lsem p\rsem(p^*,d) = p^*$$
Program $p^*$, when run, will print its own text,  regardless of what its input was.

\item \label{app-self-recogniser}
  A {\em self-recognising } program:  consider  the 
  program\footnote{We enclose program texts in boxes and use {\tt teletype} font. Reasons: to emphasise their syntactic nature, e.g., to distinguish a program from the mathematical function it computes.   In this paper programs are always imperative or  first-order functional.}
\vair

\hspace{30mm} $p$ = \fbox{\tt
          p(q,d) = if d=q then 1 else 0
}
\vair

Then  $\lsem p\rsem(p^*,d) = \lsem p^* \rsem (d)$ implies: \hair\hair
$\forall d\in D\hair \ba{lcl}\lsem p^* \rsem (d) &=& 1 \mbox{\ if\ } d = p^*\\
\lsem p^* \rsem (d) &=& 0 \mbox{\ if\ } d \neq p^*\\
\ea$
\vair

Program $p^*$, when run, will print $1$ if its input is its own text, and print $0$ otherwise.\vair
\vair

\item Further, Kleene's theorem has many  important  applications in recursion theory, e.g., see the extensive overview by 
Moschovakis \cite{moschovakis}.
Many applications (perhaps most)  require using the universal machine (a self-interpreter). 
\smallskip

\item A major consequence of Kleene's theorem (and the reason for the theorem's name) is that it implies that {\em any programming language satisfying Rogers' axioms} is {\em closed under 
recursion}.\footnote{Meaning: any function defined by recursion from programmable functions is itself programmable. See Kleene's Section 66, and the discussion around Theorems XXVI and XXVII \cite{kleenebook}.}
This is not immediately evident for, say, Turing machines since the Turing machine architecture has no recursive language constructions whatever.

Closure of Turing-machines under recursion follows from Kleene's theorem, since  this computation model satisfies Rogers' axioms. The trickiest part is the universal machine property. It was shown by Turing in \cite{turing},   using   a nonrecursive construction.

\vair

\item\label{app-recursion-removal}
  {\em Recursion removal}: For an example, consider the program
\vair
 
\bc
$p$ = \fbox{
{\tt\bt{l}
         p(q,d) = if d=0 then 1 else d * univ(q, d-1)\\
         
         univ (program, data) = ... ; universal program\\
\et}}
\ec
\vair

\fl It is immediate that
$$\ba{lcccc}
\lsem p^* \rsem (0) &=& \lsem p \rsem (p^*,0) &=& 1 \\

\lsem p^* \rsem (d+1) &=& \lsem p \rsem (p^*,d+1) &=& (d+1)* \lsem p^* \rsem(d)\\
\ea$$

\fl
Thus $  \lsem p^* \rsem(d) = d!$, the factorial function.
Net effect:  $p^*$ computes $d!$, even though {\em $p$ can be defined completely without recursion}
(e.g., {\tt univ} could be  Turing's universal program).

\item A mind-boggling application involves {\em interchanging the role of programs and data.} 
Consider the program
\vair

\bc
$p$ = \fbox{
{\tt\bt{l}
         p(q,d) = univ(d,q)\\
         
         univ (program, data) =  ... ; universal program\\
\et}}
\ec
\vair

Then the $p^*$ that exists by Kleene's theorem satisfies:
$$
 \forall\ q \in \mathit{Pgms}\  .\  \lsem p^* \rsem(q) =\lsem p \rsem(p^*,q) = \lsem q \rsem(p^*)
$$

\ee

\subsection{Kleene's proof  of SRT}
\vair

\fl The first step, given $p$, is  to find  a program $\tilde{p}$ 
such 
that
$$
\forall q  \in \mathit{Pgms}\  \forall d  \in D \ .\  \lsem \tilde{p}\rsem (q,d) = \lsem p\rsem (\lsem \spec \rsem(q,q), d)  
$$
Function $f(q,d) = \lsem p\rsem (\lsem \spec \rsem(q,q), d) $ is  computable since $p$ and $\spec $ are assumed computable. By Turing completeness, there exists a program $\tilde{p}$ to compute $f$.  

Second step: build $p^* := \lsem \spec \rsem(\tilde{p},\tilde{p})$. 
Now to show that the constructed program works correctly:  
$$
\lsem p^*\rsem(d) = 
\lsem  \lsem \spec\rsem(\tilde{p},\tilde{p})\rsem (d) =
\lsem \tilde{p}\rsem(\tilde{p},d) =
\lsem p\rsem( \lsem \spec\rsem(\tilde{p},\tilde{p}),d) =
\lsem p\rsem(p^*,d) $$ 
{\bf End proof}.  
\vair

\subsection{The Moss proof  of SRT}
\label{sec-moss-proof}

Lawrence Moss  \cite{moss} proved SRT (for a specific language   {\tt 1\#}) by reasoning similar to Kleene's, but with 2 computation stages and without   the {\em S-1-1} property. First   the reasoning; the {\tt 1\#}-specific details are deferred to Section \ref{sec-moss-language}.\vair

\fl{\bf Moss'  proof  of SRT: } The first step, given $p$, is to construct a program $\hat{q}$ such that
$$
\forall r \in \mathit{Pgms} \ \forall d \in D \ .\ \lsem \lsem \hat{q}\rsem(r) \rsem(d) = \lsem p \rsem (\lsem r\rsem(r),d)\ 
$$

\fl Section \ref{sec-moss-language}  shows that $\hat{q}$ exists in language {\tt 1\#}. (Turing completeness is not enough since $\lsem \hat{q}\rsem$ is nested.)

Second step: construct
$
p^* := \lsem  \hat{q}\rsem( \hat{q}) 
$.
Now we show that the constructed program works correctly:  
$$
\lsem p^*\rsem(d) =  
\lsem  \lsem \hat{q}\rsem( \hat{q}) \rsem (d) =  
\lsem p \rsem (  \lsem \hat{q}\rsem( \hat{q}), d) = 
\lsem p  \rsem (p^*,d)
$$
\fl{\bf End proof}.

\subsection{Remarks on constructing the ``fixpoint''}

In both constructions only $\lsem\_\rsem^1$  and  $\lsem\_\rsem^2$ are used: programs are run  with either 1 or 2 arguments. The only connection between $\lsem\_\rsem^1$  and  $\lsem\_\rsem^2$ is the {\em S-1-1} function (Kleene), or the construction of $\hat{q}$ (Moss).
The programs involved in Kleene's proof have {\em no recursion} and  {\em  no nested loops}.
(This is no surprise, since computers and  programming languages had not yet been invented when the SRT was proven.)

Finally,  the SRT proofs make {\em no use of the universal function} at all. (It is only used in applications, e.g., as in the applications above to show closure under recursion or to interchange data and program.)

\section{Towards computer realisations of the  SRT proof }

The classical  G\"odel number-based constructions well-known from recursive function theory (Kleene, Rogers, Gandy\cite{kleenebook,rogers,gandy}) are quite impractical to implement, as the techniques are based on numerical encoding: prime power exponentiations  and factoring. Numerical encoding was reasonable for their purpose, which was   to delineate the boundaries of computability and not to study  complexity. We wish, however, to see how to implement  such constructions efficiently  on a computer.

A critical step in Kleene's proof is going from the {\em mathematical function} $f$ to the {\em program} $\tilde{p}$.
By appealing to  Turing completeness, the Kleene proof  avoids being tied to any one programming language. 
Our goals are different, and to talk about SRT complexity, we will need both {\em more concrete 
computation models} and {\em explicit program constructions}.

\subsection{The ``Swiss pocket knife''}
\label{sec-swiss-pocket-knife}

The imperative flow chart language of \cite{JGS} is enough to carry out all the SRT applications above. 
We will see, as did  Bonfante and Greenbaum  in \cite{bonfantegreenbaum}, that {\em the constructions to prove the SRT} can be done in a very small subset   {\sc tiny} of the
flow chart language of \cite{JGS} such  that {\em all programs run in constant time}.\footnote{The  {\sc tiny} language is of course not Turing-complete. In brief, the Turing-complete imperative languages of \cite{stoc,Jones:97:ComputabilityComplexity,JGS} are in essence   {\sc tiny} plus  {\tt while} and  {\tt if} commands.}

\subsection{Programs as data}

Programs have been formulated  in computability theory in many ways, e.g., as
a natural number (a G\"odel number, by Kleene and others \cite{kleenebook,kleene,rogers}); a Turing machine program or set of quintuples
\cite{turing};
a lambda expression by Church \cite{church}; a set of recursive function definitions  (by Kleene and others \cite{kleenebook}); a set of rewrite rules or a register machine (see Minsky in  \cite{minsky});
 an S-expression in McCarthy's {\sc lisp}  \cite{mccarthy}), and many others.
In classical recursive function theory a program is a natural number from $\nats$.

A  programming language view is that a program is  {\em  an abstract syntax tree} (e.g., an  S-expression
as in {\sc scheme} or {\sc lisp}).
Advantages: abstract syntax trees such as S-expressions give more natural versions of
the {\em size} $ |p|$ or $|d| $ of a program $p$ or a data 
value $d$. 
This is important because the relation between  input size and program running time is central in computer science,
cf.\ the {\sc p}={\sc np} problem.

More accurately, the semantics of {\sc lisp}-like languages are not really based  on trees, but rather on {\em DAGs} (directed acyclic graphs), since substructures of data may be (and usually are) 
shared. This can be critical for measuring running 
times.\footnote{Tree-based and DAG-based models define the same input-output relations for programs without selective updating such as {\tt set-car!}.}

Amtoft et al.   \cite{botik} tried  Kleene's SRT construction in a first-order {\sc lisp}-like functional programming language 
with tree-structured data, encountered problems, and modified the language. See Section \ref{sec-botik-overview}.

\subsection{The Kleene SRT proof with tree-structured data }
\label{sec-tree-structured-data}

To make Kleene's  construction  computationally explicit one can use  imperative flow chart programs with {\sc lisp}-like data 
as in \cite{JGS} Chapter 4. A very small subset suffices for this paper: 
the {\sc tiny} language of Bonfante and Greenbaum \cite{bonfantegreenbaum}.
Program  format:
$p$ = \fbox{{\tt read x1,...,xn; C; write out}} with $n \geq 0$.
Here   {\tt C} is a command built from assignments {\tt x := e}, 
sequencing {\tt C1;C2}  and 
expressions with 
variables, constants {\tt 'd}, 
and operators {\tt hd}, {\tt tl}, {\tt cons}.
There are no tests or loops, so {\em every program   will run in constant time} 
(assuming as usual in DAG-based semantics that operators {\tt hd}, {\tt tl}, {\tt cons} are constant-time).

\paragraph{Program specialisation:} 
Let program $p$ = \fbox{{\tt read q, d; C; write out}}, and let $s$ be a ``static''  value for  variable {\tt q}. 
The result of  specialisation could be
$$
p' = \fbox{\mbox{\tt read d; q := 's; C; write out}}
$$
This specialisation gets $p'$ by removing from $p$ the input of variable {\tt q}, and adding  assignment {\tt q\ := 's}. It  inserts the static data value {\tt s}   {\em inside} the constant {\tt 's}.
More generally, we need a concrete program $\spec$ such that 
$$\forall pgm   \ \forall s,d\ .\ 
 \lsem pgm \rsem (s,d) = \lsem \lsem \spec \rsem (pgm,s) \rsem (d)
 $$
 The form of the specialiser is
$$
\spec = \fbox{\mbox{\tt 
read\ pgm, s; \ C$_{spec}$; \ write\ outpgm}}
$$
where command {\tt C}$_{spec}$ is the ``body'' of $\spec$. 
To proceed further we need to be more specific about the form of programs as data:
concrete tree structures to represent 
$p, p' $and $\spec$.
Following the lines of \cite{stoc,Jones:97:ComputabilityComplexity,bonfantegreenbaum} we can use 
a {\sc lisp}-inspired {\em concrete syntax}, e.g.,
$$
p = {\tt ((q\ d)\ \ \underline{C}\  \ out)}
$$
where {\tt \underline{C}} is a ``Cambridge Polish'' representation of 
{\tt C}.\footnote{ 
\underline{\tt 'd} = {\tt (QUOTE d)}, 
\underline{\tt op e} = {\tt (op \underline{e})},  
\underline{\tt cons(e,e')} =  {\tt (cons \underline{e} \underline{e'})},
\underline{\tt x := e} = {\tt (:= x \underline{e})}, and
\underline{\tt C1;C2} = {\tt (;  \underline{C1} \underline{C2})}.}
Representation of specialisation result   $p'$   above:
$$ 
p' = {\tt ((d)\ \ (;\  (:=\  q\  (QUOTE\  s))\ \  \underline{C})\  \ out)}
$$
Obtain $\lsem\spec\rsem(p,s) = p'$ by defining the
 body {\tt C$_{spec}$} of $\spec$ to be (in {\sc lisp}-like informal 
syntax\footnote{More notation: 
{\tt list(e1, e2,..., en)} is short for
{\tt cons(e1, cons(e2,..., cons(en, '())...))}.}):
\vair

\fl$C_{spec} = $  
{\tt \fbox{
\bt{ll}
inputvar&:=\ hd hd pgm; C\ :=\ hd tl pgm; outputvar :=\ hd tl tl pgm; \\
initialise&:=\ list(':=, inputvar, list('QUOTE, s)\ ); \\
body&:=\ list('; ,initialise, C\ ); \\
outpgm&:=\ list(tl hd pgm, body, outputvar);
\et}}
\vair

\subsubsection{Program details for the Kleene construction with tree-structured data}
\vair

\be

\item Let $p=$ {\tt \fbox{read\ q,d;\ C$_p$;\ write\ out}} and let {\tt s} be the known value for {\tt q}.
Let {\tt C$_{spec}$} be the body of specialiser $\spec$. As above  $\lsem \spec\rsem(p,s) = p' = \ {\tt \fbox{{\tt read\ d;\ q\ := 's;\ C$_{p}$;\ write\ out} }}$.

\item\label{buildr}  Build $ \tilde{p} =$   
{\tt \fbox{
\bt{l}
read\ q,d;\\
pgm\ :=\ q;\ s\ :=\ q;\ C$_{spec}$;  \ (* specialise q to q *)\\
q \ :=\ outpgm;\ C$_p$; \hspace{12mm} (* then run p on the result *)\\
write\ out 
\et}}
\vair 

\fl This  clearly satisfies $ \ \lsem \tilde{p}\rsem(q,x) = \lsem p\rsem( \lsem \spec\rsem(q,q),x)$. 
\vair

\item\label{buildpstar} Let $ p^* =  \lsem \spec\rsem(\tilde{p},\tilde{p}) =$   
{\tt \fbox{
\bt{l}
read\ d;\\
q := '$\tilde{p}$;  \hspace{24mm}   (* initialise q  to program $\tilde{p}$ *)\\
pgm\ :=\ q;\ s\ :=\ q;\ C$_{spec}$;\  (* specialise $\tilde{p}$ to $\tilde{p}$ *) \\
q \ :=\ outpgm;\ C$_p$; \hspace{6mm} (* run p on the result *) \\
write\ out 
\et}}
\vair

\fl This satisfies $\lsem p^*\rsem(d) = 
\lsem  \lsem \spec\rsem(\tilde{p},\tilde{p})\rsem (d) =
\lsem \tilde{p}\rsem(\tilde{p},d) =
\lsem p\rsem( \lsem \spec\rsem(\tilde{p},\tilde{p}),d) =
\lsem p\rsem(p^*,d)$. 

\ee

\paragraph{Program self-reproduction:}  
The start 
\fbox{{\tt  q := '$\tilde{p}$;  pgm :=\ q;\ s\ :=\ q;\ C$_{spec}$;\  q  := outpgm}}
of the  $p^*$ program 
  assigns to {\tt q}  the value $ \lsem \spec\rsem(\tilde{p},\tilde{p})$, which equals $p^*$. The net effect is that this code segment in $p^*$ assigns to {\tt q}   {\em the text of the entire   program $p^*$  that contains it}.
  
\subsubsection{Constant time, and the role of shared data-structures}  
\label{sec-shared-data-structures}

It may  be surprising that Kleene's SRT can be proven by such simple means. 
{\sc tiny} is a very limited language, since a {\sc tiny} program can only access (by means of {\tt hd}, {\tt tl}) parts of its input that lie a fixed distance from the root of its input; the program  is indifferent to the remainder of its input. This implies that  every 
{\sc tiny} program runs in constant time. 

An analysis of the size of $p^* = 
\lsem \spec\rsem(\tilde{p},\tilde{p})$: program $p^*$ contains 
\bi
\item a copy of  the body {\tt C$_{p}$} of program $p$, and a copy of  the body {\tt C$_{spec}$} of the specialiser; plus

\item a copy  of program $\tilde{p}$ (in   {\tt '}$ \tilde{p} =$ {\tt (QUOTE $\tilde{p}$)}).
This $\tilde{p}$ {\em also contains copies} of both  {\tt C$_{p}$} and  {\tt C$_{spec}$}.
\ei
These copies are shared in the natural implementaion: variables {\tt pgm} and {\tt s} in  programs $\tilde{p}$ and $p^*$ all refer to the same DAG node. 
One effect is that  a printed-out version of $p^*$ may be considerably larger than $p^*$ as a DAG, beause of the  shared substructures.

\subsection{A reflective extension of the programming language}
\label{sec-botik-overview}

Amtoft et al \cite{botik} observed a practical problem in the Kleene construction in the case that program $p$ calls the universal program $univ$. Implementing recursion 
as in Application \ref{app-recursion-removal} gave a surprise:  in order to compute $n!$  the self-interpreter is applied  {\em to interpret itself} at $n$ meta-levels. {\em Consequence}: when applied to compute $n!$ the Kleene construction {\em takes exponential time}, and not linear time as one might expect. 
(Remark: the Moss construction would have the same problem.)

A  design change: the functional language of  \cite{botik} was given a ``reflective extension''.
First, a new constant {\tt *} was added to the language. Its value: {\em the text of the program currently being executed}. 
Second, a new call form {\tt univ p d} was added, yielding value $\lsem {\tt p}\rsem({\tt d})$.
With the aid of these  new constructions it was straightforward to construct a program $p^*$ as needed for the Kleene result, and without self-application. The resulting $p^*$ evaluated $n!$ in linear time, albeit with a significant interpretation overhead.

The rationale behind this perhaps unexpected language design was that an interpreter was being used to execute programs. Since the interpreter always has to  have  the program it is interpreting at hand,  the value of constant {\tt *} is always available. Further, a source program call  {\tt univ p d} can be implemented by a simple recursive call to the currently running interpreter (thus sidestepping the need to  interpret an interpreter, etc.).

Conclusions:  the construction of \cite{botik}  gives a  more efficient output program than Kleene's version;  
but the  ``reflective extension''  is somewhat inelegant (even hacky); and an efficiency drawback is that {\em every program execution} involves a significant interpretation overhead.

\subsection{The  Moss SRT proof with {\tt 1\#}}
\label{sec-moss-language}

The Moss approach  to construct $\hat{q}$ from Section \ref{sec-moss-proof} recapitulated: 
The  language {\tt 1\#} is based on  {\em term register machines} (TRM for short): a  variant of  Shepherdson and Sturgis' well-known register machines  \cite{minsky,shepherdsonsturgis}, generalised
to work on strings from $D = \{1,\#\}^*$  as data instead of natural 
numbers. 
A program operates on a fixed number of registers $R1, R2,\ldots, Rk$. To compute $\lsem p\rsem^n$, the program inputs are in Registers $R1,R2,\ldots, Rn$, and output is  in Register $R1$.
Language {\tt 1\#} is acceptable   since it is possible to construct a universal program and programs  for the {\em S-1-1} functions.

A {\tt 1\#} program, as well as the data it operates on, is a string from $\{1,\#\}^*$.  
For  program representation details, see \cite{moss}. 
A key point is that there exists  a {\em program composition operation} $|$ for  {\tt 1\#} such that
$$
\forall p, q\in \mathit{Pgms}\ \forall x \in D\ .\ 
\lsem\ p\ |\  q \ \rsem(x) = \lsem q\rsem(\lsem p\rsem (x))
$$
Operation $|$ is just ``append'', i.e., string concatenation.
Further, there exist terminating programs 
${\tt move}_{i,j}$ and {\tt write}, {\tt diag} as follows.  Their {\tt 1\#} codes are also in \cite{moss}, all using 3 or fewer registers. 
\be
\item ${\tt move}_{i,j}$ appends the contents of $Ri$ to the right end of $Rj$ (and empties $Ri$ in the process).
\item For any $x \in D$,
$ \lsem\ \lsem {\tt write} \rsem (x)\  \rsem() = x$.
\item For any $r \in \mathit{Pgm}$,
$\lsem\ \lsem {\tt diag} \rsem (r) \ \rsem() = \lsem r \rsem(r)
$.\ee
{\em Effects}: program ${\tt write}$ produces from input  string $x$ a program that, when run,  writes $x$.
Program $ {\tt diag}$  produces from input   $r$ a program that, when run, computes $ \lsem r \rsem(r)$.
Conceptually, ${\tt write}$ expresses the essence of {\em code generation}; and
$ {\tt diag}$  expresses the essence of {\em self-application}.\footnote{An example of a self-reproducing program
similar to  Application \ref{app-self-reproducer}
is easy to construct directly:
Define
$
\mathit{{\tt self}} := \lsem {\tt diag}\rsem ({\tt diag})
$.
Then
$
\lsem \mathit{{\tt self}} \rsem () = \lsem\ \lsem {\tt diag} \rsem ({\tt diag})\  \rsem() 
= \lsem {\tt diag} \rsem ({\tt diag}) = \mathit{{\tt self}}
$.}
\vair

\fl The Moss construction explicitly builds a program satisfying the requirements of  Kleene's proof.  
The first step in in the Moss SRT construction: given $p$, construct 
$$
\hat{\tt q} = 
{\tt diag} \ |\  
{\tt move}_{1,2} \ |\  
\lsem{\tt write}\rsem(\mv{1}{4}) 
\ |\  \mv{2}{1} \ |\  
\lsem{\tt write}\rsem(\mv{4}{2})  \ |\  
\lsem{\tt write}\rsem({\tt p}) 
$$
Program $\hat{\tt q}$ is  terminating since all of its parts are terminating. 
Given the properties of program composition and the {\tt write}, {\tt diag} and the $\mv{i}{j}$ programs, it is easy to see that
$$
\lsem \hat{q} \rsem(r) = 
\mv{1}{4} \ |\  
\lsem{\tt diag}\rsem(r ) \ |\  
\mv{4}{2}
  \ |\  {\tt p}
$$
which implies
$
\lsem \lsem \hat{\tt q} \rsem(r) \rsem(d) = \lsem p \rsem (\lsem r \rsem(r),d)\ 
$ as required.
The second step is to set $p^* = \lsem\hat{q}\rsem(\hat{q})$.

\section{Operational questions about theoretical constructions}
\label{sec-operational-questions-theoretical-constructions}

\subsection{Program running times}

Write $\mathit{time}_p (d)$ for the number of  steps to compute $\lsem p \rsem (d)$ in a suitable computation model, e.g., a programming language. Assume  given a
function $\mathit{time}_p (d)$ that satisfies the Blum {\em machine-independent complexity axioms}  
\cite{blum,machteyyoung}:

\be
\item For any $  p \in \mathit{Pgms}, d \in D$,  $\mathit{time}_p (d)$  terminates iff
$ \lsem p \rsem (d)$ terminates.

\item The property $\mathit{time}_p (d) \leq t$ is decidable, given program $p$, input data $d$ and time $t$.

\ee
For instance in an imperative language one could count $1$ for each executed assignment {\tt :=},   operator,  and variable or  constant access.

\subsection{Some natural questions for a computer scientist}  
\label{sec-self-time}

\be

\item What is the {\em computational overhead} of self-interpretation, i.e., applying a universal machine ?

\item Can specialisation as in the {\em S-1-1} axiom {\em speed a program up}? If so, by how much?

\item {\em How hard to construct} are the programs that exist  by Kleene's second recursion theorem;
and {\em how efficient} are they (or can they be)?

\ee
   Questions 1 and 2 were motivated in 1971 by Futamura  (reprinted in 1999 \cite{futamura}); some answers are given by Jones, Gomard, Sestoft in  \cite{JGS}.
For context, first a brief review  1 and 2 from the viewpoint of  \cite{JGS,futamura}. 
Following this, we  obtain some new results about question 3 (investigated earlier in \cite{botik,moss}).

\subsection{Interpretation overhead} 

By the Rogers axiom,
$\lsem p\rsem(d)$ terminates iff 
$\lsem univ\rsem(p,d)$ terminates. 
By the first Blum complexity axiom 
$$
\forall\ p\  \forall d\  (\exists t \ .\ \mathit{time}_p (d) \leq t) \mbox{\ iff\ }  (\exists t' \ .\ \mathit{time}_{univ} (p,d) \leq t')
$$
Interpretation overhead is the efficiency slowdown caused by use of a universal machine, i.e., the
relation between (the smallest such) $t$ and $t'$.
Their existence does not, however, imply there is any simple relation between them. 
Some possibilities for interpretation overhead:

\be

\item $\forall p\  \forall d\  \exists c \ .\   \mathit{time}_{univ} (p,d) \leq c \cdot \mathit{time}_p (d) $
\hfill (always true)

\item $\forall p\  \exists c  \  \forall d\ .\ \mathit{time}_{univ} (p,d) \leq c \cdot \mathit{time}_p (d)   $
\hfill (program-dependent overhead)

\item $ \exists c \  \forall p\  \forall d\ \ .\  \mathit{time}_{univ} (p,d) \leq c \cdot \mathit{time}_p (d) $
\hfill (program-independent overhead)

\ee
One might expect Overhead 2 in practice, reasoning that
if an interpreter $univ$ simulates $p$ one  step at a time, then $t' \leq f(p) \cdot t$ for some function $f$. If so, 
then the  interpretation overhead may depend  on the program being interpreted, but not on the current input data $d$.

Unfortunately this  is not always so.
One counterexample is Turing's original universal machine \cite{turing}. Because of the 1-dimensional tape, simulation of the effect of one quintuple in program  $p$ may require that the interpreter scans {\em from} the tape area where $p$'s program code is written, {\em to} the area where $p$'s currently scanned data square is found, and then {\em scans back again} to  $p$'s program code. Worst-case:  $\mathit{time}_{univ}(p,d)$ is  larger than  $p$'s running time multiplied by the  entire size of of its run-time
data area. 

The same problem appears in most published universal machines, including the TRM model. The problem is  the need to ``pack'' all  the simulated program $p$'s data values into one of $univ$'s data values. Applied to TRMs: although $univ$ has only a fixed number of registers, there exists no limit to the number of registers that an interpreted program $p$ may have.
The root of this problem is that {\em a limit is inherited from the interpreter},  e.g., the number of registers. Mogensen describes this problem of inherited limits  in general terms and with many specific instances  in \cite{inheritedlimits}.

{\em Is this a  problem?} 
Yes (from this paper's viewpoint)  since a self-interpreter is needed 
for most applications of the recursion theorem 
(beyond self-reproduction and self-recognition).
\vair

\fl Smaller overheads have been obtained for some computation models. 
Interpretation overhead  2 is typical for interpreters with tree-structured data and constant-time pointer access, e.g., interpreters expressed in {\sc scheme}, {\sc prolog}, etc.,  and the $\lambda$-calculus. The partial evaluation literature (overviewed in \cite{JGS}) contains many such self-interpreters.  Mogensen \cite{DBLP:journals/lisp/Mogensen00} has detailed analyses of the costs of  $\lambda$-calculus self-interpretation under several execution models. 

Overhead 3 is seen in  \cite{stoc} for a very limited  language (with one-atom trees and one-variable programs); and for the $\lambda$-calculus using some of Mogensen's models and cost measures  \cite{DBLP:journals/lisp/Mogensen00}. 
The first assumes constant-time pointer access, and the second assumes constant-time variable access  (or does not count it). Overhead 3 also holds for the biologically motivated ``blob'' computation model \cite{blob} which has 2-way  bonds, no variables, 
bounded  ``fan-in'' among data values,  and a single 2-way activation bond between program and data (which must always be adjacent).

\subsection{Futamura projections: partial evaluation can remove interpretation overhead
}

Partial evaluation concerns {\em efficient implementation} of the  {\em S-m-n} property.
A partial evaluator is simply an {\em S-1-1}  program $\spec$ as in the Rogers axioms.
Supposing $univ$ is a universal program,
the following properties (due to Futamura 1971
\cite{futamura}) are easy to verify from the definition of   $univ$ and $\spec$. The first line asserts that 
a partial evaluator can compile a program $\mathit{source}$ into a semantically equivalent  program $target$. The second line says that 
a partial evaluator can generate a compiler; and the third, that 
a partial evaluator can generate a compiler generator. 

\vair

$
\ba{|llclc||cll|}   \hline
& \multicolumn{3}{c}{\mbox{\bf Definitions}} &&  \multicolumn{3}{c|}{\mbox{\bf Properties}} \\ \hline
1. & \hair target &:=& \lsem \spec \rsem (univ, \mathit{source})  && \hair \forall d\   .\ \lsem target \rsem (d) &=& \lsem \mathit{source} \rsem (d)\ \\
2. &\hair \mathit{compiler} &:=& \lsem \spec \rsem (\spec, univ)   &&  \hair target &=& \lsem \mathit{compiler} \rsem (\mathit{source})\\
3. & \hair cogen &:=& \lsem \spec \rsem (\spec, \spec)   && \hair \mathit{compiler} &=& \lsem cogen \rsem (univ)\\ \hline
\ea
$
\vair

\fl The Futamura projections   involve
program self-application, but in a way different than that used in the proof of Kleene's theorem, e.g.,
$\lsem \spec \rsem (\spec, univ)$ rather than $\lsem \spec \rsem(q,q)$ or $\lsem \hat{q} \rsem(\hat{q})$. 

The Futamura projections were first fully realised on the computer in Copenhagen in 1985; see \cite{JGS} for details and references.
The expensive self-application in the table  for $\mathit{compiler} := \lsem \spec \rsem (\spec, univ)$
can be avoided by doing  another computation that only needs doing once:  
$cogen := \lsem \spec \rsem (\spec, \spec)$. After that, an individual compiler can be generated from any interpreter $int$ by the significantly faster run $\mathit{compiler} := \lsem cogen \rsem (int)$. 
For details, see \cite{JGS}.
A moral: one deep self-application, to construct $cogen := \lsem \spec \rsem (\spec, \spec)$, can be used in place of  single  self-applications  $\mathit{compiler} := \lsem \spec \rsem (\spec, int)$ that are done repeatedly.

We hope that such analogies will  lead to a better complexity-theoretic understanding of Kleene's second recursion theorem.
\vair

\fl Complexity issues in partial evaluation are fairly well-understood and partial evaluators  well-engineered. 
An example ``optimality'' result from \cite{JGS}, for a simple first-order {\sc scheme}-like language: 
\vair

\fl {\bf Theorem} Partial evaluation can remove {\em all interpretation overhead}, meaning  
$$ \mathit{time}_{target}(d) \leq  \mathit{time}_{\mathit{source}}(d)$$ 
for all data $d$ and a natural self-interpreter $univ$.
\vair

\fl The removal of all interpretation overhead has been achieved in practice as well as in theory.

\section{Operational aspects of Kleene's second recursion theorem}
\label{sec-kleene-constructive}

In spite of the theorem's high impact on theory, it is not easy to reason about its efficiency, 
e.g., time usage. To being with, there are two distinct efficiency questions with rather different answers:
\bi
\item The time  it takes {\em to construct $p^*$ from $p$}; 
and
\item {\em The efficiency of the constructed program} $p^*$, when run on $d$.
\ei

\fl The constructions used to prove Kleene's theorem are not  complex,
and do not require the full power of recursion theory to construct program $p^*$ from  
$p$.\footnote{The complexity of running $ p^*$, i.e.., of computing  $\lsem p^*\rsem(d)$, can, however be high, depending on program $p$.}

\subsection{Some operational detail} 
\label{some-operational-detail}

The  Moss approach  is similar to Kleene's, but with different ``building blocks,'' e.g., no 
{\em S-m-n} theorem is used.  Based on computer experiments:
\be

\item In practice the  Moss approach is somewhat faster and simpler than the Kleene approach, but 
the transformed program $p^*$ works in essentially  the same way for both constructions.

\item How $p^*$ works: for a given $p$,
\bi
\item $p^*$ first computes {\em a copy of itself, including $p$}.

\item It then runs $p$ on the copy of  $p^*$, together with the data input $d$. 
\ei

\item  When run, program $p^*$ has been observed  to be {\em large and slow}. 
Both constructions generate a {\em rather expensive set-up phase}, to make the copy of $p^*$,
 before ever looking at $p$'s data input $d$.

\item The generated program $p^*$  may contain {\em more than one version of $p$ and $\spec$}, in plain and code-generating versions. 
(This has already been seen in Section \ref{sec-shared-data-structures}.)

\ee

\subsection{Corner cases} 

Three  ``corner cases''  (a term due to Polya) that may give some insight into operational behavior:
\be

\item \label{corner-case-projection1}
First projection: $\lsem p\rsem(e,d) = e$. In this case $p^*$ is a self-reproducing program, as in Application \ref{app-self-reproducer}.

\item \label{corner-case-projection2}
 Second projection: $\lsem p\rsem(e,d) = d$. By SRT  $\lsem p^*\rsem(d) = \lsem p\rsem(p^*,d) = d$, so $p^*$ computes the identity function, but slowly.  It first constructs a copy of itself and then runs $p$, ignoring the copy  it made.

\item  \label{corner-case-univ}
$p = \mathit{univ}$. By SRT  $\lsem p^*\rsem(d) = \lsem p\rsem(p^*,d)= \lsem univ\rsem(p^*,d) = \lsem p^*\rsem(d)$, which makes no restriction at all on $p^*$. The resulting program  $p^*$ loops infinitely.

\ee

\subsection{Can more efficient SRT output  programs be obtained?} 

In the special case that $p$ {\em does not  use its first argument} $p^*$, as in in corner  case  \ref{corner-case-projection2}, one would expect
\bi
\item $
\mathit{time}_{p} (p^*,d) = f(d)$ for some function $f$

\item  $
\mathit{time}_{p^*} (d) = c + f(d)$ for some constant $c$ (cf. Section \ref{some-operational-detail})
\ei
These expectations hold in computer experiments, but the constant $c$ is {\em very large}.

The unexpected exponential time behavior of the factorial example 
in application \ref{app-recursion-removal} could be circumvented as in  Section \ref{sec-botik-overview} and \cite{botik}, but at considerable cost: {\em interpretive execution of all programs}. Can this effect be achieved more economically, e.g., by a stronger $\spec$ algorithm?

\subsection{Utility of a more efficient program specialiser.} Kleene's proof is based on the {\em S-m-n} construction, so would be  natural to expect the Kleene SRT construction to benefit from using a state-of-the-art partial evaluator, e.g., as described in \cite{JGS}. 

Bonfante (continuing the line of \cite{bonfantegreenbaum}) added  to the end of $\spec$ a simple optimiser:  a ``dead code'' detector and eliminator. This was enough to eliminate all the unnecessary computation seen in corner  case \ref{corner-case-projection2}.
It is less clear, however, where such optimisations could be put into the Moss construction.

\subsection{Relating the  {\tt 1\#} and {\sc tiny} SRT constructions} 
\label{sec-relating-1sharp-and-tiny}

\subsubsection{Some experiments with {\tt 1\#}.} 

The results reported by Moss in \cite{moss} led this paper's author to develop a straightforward {\tt 1\#} implementation in {\sc scheme},
with a step counter to evaluate running times. Some comments:
\bi
\item {\tt 1\#} is a register machine model, so program {\tt move}$_{i,j}$ (used to  assign {\tt R$_i$:=R$_j$}) takes time $O(|R_j|)$. 

\item Data structures are the main difference between {\tt 1\#} and {\sc tiny}.  The linear strings in $\{{\tt 1},{\tt \#}\}^*$  must be scanned one bit at a time, in contrast to {\sc tiny}'s constant-time pointer operations. 

\item Observed for the Moss   SRT construction:
\bi
\item Computing $\lsem {\tt write}\rsem(x)$  takes time $O(|x|)$, as does $\lsem {\tt diag}\rsem(x)$. 

\item For a  small $p$, the set-up phase of $p^*$ (computing $\lsem\hat{q} \rsem (\hat{q})$) takes between 20,000 and 40,000 steps (depending on implementation choices). 

\item A significant factor in computing $\lsem\hat{q} \rsem (\hat{q})$ was the time  $\lsem {\tt write}\rsem(x)$ and $\lsem {\tt diag}\rsem(x)$ used to read $x$, and to compute their output values while  scanning several versions of program $p$.

\ei

\ei
We also implemented the Kleene version of the SRT construction in  {\tt 1\#} (specialising by {\tt s}$^1_1$ as in \cite{moss}). It ran about twice as slowly as the Moss SRT construction.
Further, experiments were  done with a  {\tt 1\#} variant with constant-time assignments; this is natural for programming languages. The resulting Moss SRT construction  ran roughly twice as fast as the original {\tt 1\#}. \vair

\subsubsection{The Moss construction in constant time using {\sc tiny}.}

Every {\sc tiny} program runs in constant time independent of the size of its input, including the Kleene SRT construction  seen in Section \ref{sec-tree-structured-data}
We will not re-do the complete proof of Section \ref{sec-moss-language}, but just show that central components of   the Moss construction are expressible in {\sc tiny}.

First, program composition $|$ : Let 
$p= $ \fbox{\tt read x; C$_p$; write y} and 
$q= $ \fbox{\tt read y; C$_q$; write z}.  
Without loss of generality, $p,q$ have disjoint variables, except for {\tt y}.
{\sc tiny} has no need for a time-consuming ``append'' operation, since
program  
$$p \ |\ q =  \mbox{\fbox{\tt read x; (C$_p$ ; C$_q$); write z}} 
$$ behaves as required, satisfying $\lsem p \ | \ q \rsem(x) = \lsem q\rsem ( \lsem p \rsem(x)$. 
Expressed in concrete syntax, the program composer should transform  program inputs 
{\tt ((x) \underline{C$_p$} y)} and 
{\tt ((y) \underline{C$_q$} z)}
into
{\tt ((x) (; \underline{C$_p$} \underline{C$_q$}) z)}.
This is straightforward to program in {\sc tiny}.

Next, we need a program such that $\lsem\lsem {\tt write}\rsem(x)\rsem() = x$  for any $x \in D$. 
For example, $\lsem {\tt write}\rsem(x)$ could yield as output the {\sc tiny} program 
$w_x =  $ \fbox{\tt read; out := 'x; write out} . In concrete syntax:
\bc
$w_x = $ {\tt (() (:= out (QUOTE $x$) out)}
\ec
A program {\tt write} to generate $w_x$  from $x$:
$$
{\tt write} = \fbox{\tt
\bt{ll}
read x; \\
out := list('(), (list ':=, 'out, list ('QUOTE, x)), 'out); \\
write out
\et
}
$$
A program {\tt diag}   satisfying $\lsem\lsem {\tt diag}\rsem(r)\rsem() = \lsem r \rsem(r)$,  for  $r \in \mathit{Pgms}$. Goal: $\lsem {\tt diag}\rsem(r)$ is a program
$d_r$ such that 
$\lsem d_r\rsem() =\lsem r\rsem(r)$. Let $r =$  \fbox{\tt read x; C$_r$; write out} have concrete syntax
{\tt ((x) \underline{C$_r$} out)}. 
Then $d_r$ could be
\bc
$d_r =$ \fbox{\tt read; x := 'r; C$_r$; write out}.
\ec
or, in concrete syntax:
\bc
$d_r = $ {\tt (() (; (:= x (QUOTE r)) \underline{C$_r$}) out)}
\ec
A program {\tt diag} to generate $d_r$  from $r$:
$$
{\tt diag} =  
\fbox{\tt 
\bt{ll}
read r; \\
inputvar&:=\ \ \ hd hd r; C\ :=\ hd tl r; outputvar :=\ hd tl tl r; \\
initialise&:=\ \ \ list(':=, inputvar, list('QUOTE, r)\ ); \\
body&:=\ \ \  list('; ,initialise, C); \\
outpgm&:=\ \ \  list(tl hd pgm, body, outputvar); \\
write outpgm
\et}$$
This {\tt diag}   is  just {\sc tiny} specialiser $\spec$, modified to generate code that first copies (a pointer to) static data  input {\tt r}, to build {\tt x := 'r}; followed by its  body {\tt C$_r$}. The generated code's net effect is to execute program $r$ on input $r$.  
 
 We omit the similar but tedious details of building a {\sc tiny} version of program $\hat{q}$. The ``append'' effect of $\mv{i}{j}$ is achieved in constant time by  {\sc tiny}'s ``{\tt ;}'', as used   above for $|$.

\vair

\section{Related work, future work,  and acknowledgements}

\subsection{Related work} 

Kleene's second recursion theorem  attracted interest since first published in 1938, shortly after Turing's pathbreaking 1936 work that founded computability theory \cite{turing}. 
Kleene's apparently quite theoretical result has shown a  staying power
in areas far beyond its frequent usage by recursion theorists, as well-documented by Moschovakis in 2010 \cite{moschovakis}.
One reason for such widespread interest is  the way it is proven: in essence Kleene constructed a self-reproducing program. 
This is particularly surprising since Kleene did this in a quite constructive way in the 1930s,  long before the first computer was built. 

Computer scientists have for many years repeatedly re-discovered and re-solved the goal of building  self-reproducing programs, cf. an elegant example  by Thompson \cite{thompson}.
John Case  applied this fascinating theorem in  both recursive function theory and in computer learning, e.g., see  \cite{case}.
A group at Nancy led by Jean-Yves Marion has related Kleene's theorem to  computer viruses \cite{marion}, and devised {\sc tiny}.

Our own early interest in the theory-practice interface led to a 1989 paper \cite{botik}. While that solution  worked, a drawback was that its usage  always involved at least one level of interpretation overhead. 

Since then the 2006 work by Lawrence Moss   \cite{moss} brought Kleene's result strongly to the attention of computer scientists.
Further, Oleg Kiselyov \cite{kiselyov} has recently worked on the problem from a  functional programming viewpoint, a starting point being an unusually efficient self-interpreter written in the $\lambda$-calculus  using a higher order representation of program syntax.  
The work in this paper involves much lower-level languages than the ones used by  Kiselyov and Mogensen \cite{kiselyov,DBLP:journals/lisp/Mogensen00}.

\subsection{Future work}  

This paper's results concern mostly the time to produce $p^*$ by the Kleene and Moss constructions, but very little has been said about the runtime efficiency of computing $\lsem p^* \rsem(d)$ for Turing-complete languages 
(beyond mentioning that this question   motivated \cite{botik}). 
Here it becomes interesting, since straightforwardly applying the Kleene or Moss constructions often gives 
{\em unnaturally inefficient solutions}, e.g., the nested self-interpretations  seen in the recursion example of Application \ref{app-recursion-removal}. 

Following are some questions and goals for future work.

\be

\item A  question concerning the Moss 2-stage SRT construction: can an efficient specialiser (e.g., as in \cite{JGS}) be usefully applied to the result of {\tt diag} from Section \ref{sec-relating-1sharp-and-tiny}?

\item To what extent can an efficient specialiser be used to produce better fixpoint programs $p^*$?

\item It is natural to ask  whether  nested self-interpretation can be avoided without adding an interpretive overhead to every program execution (as seen in \cite{botik}).

\item Investigate the changes to the SRT efficiency results if one supposes the implementation language has a self-interpreter {\em with only additive overhead}, meaning:
$$
 \forall p\ \exists c\ \forall d \ .\ {\mathit time}_{univ}(p,d) \leq c + {\mathit time}_{p}(d)
$$
This can be done if (1) programs are data; and (2) one has an instruction with the effect of ``goto data''. An interpreter $univ$ could first compile its input program $p$ to the language in which the interpreter itself is written; and then  jump, i.e., transfer control, to this code's first instruction.
Such effects  can be achieved in the von Neumann computation model (although finite),  in Marion's SRM model extending the
Moss {\tt 1\#} language \cite{marionroyalsociety}, and in a planned modest extension of the Blob model \cite{blob}.

\item In a Turing-complete language, running times may of course be much larger than the constant-time bounds of {\sc tiny}. If program $p$ may contain control loops, challenging problems include:  
\bi
\item how to find bounds on  $\mathit{time}_{p^*}(d)$ as built by existing constructions; and 

\item how to achieve better running times by new constructions.

\ei
These questions could be approached pragmatically, or computation-theoretically.

\item A more general problem: find relations between the self-application from Kleene's SRT, e.g., $\lsem \spec \rsem(q,q)$ or $\lsem \hat{q} \rsem(\hat{q})$; and the self-application used in the Futamura projections, e.g.,

$\lsem \spec \rsem (\spec, univ)$ or $\lsem \spec \rsem (\spec, \spec)$. 

\item Finally, suppose $p$ is ``extensional'', as in Rice's Theorem: the output value $\lsem p\rsem(e,x)$ depends only on the semantics of argument $e$, so $\lsem e\rsem = \lsem e'\rsem$ implies $\forall x\ \lsem p\rsem(e,x) = \lsem p\rsem(e',x)$.
Operationally, all that can be done with $p$'s program argument $e$ is to run it (perhaps nested, e.g., 
$\lsem e\rsem(\lsem e \rsem(x))$.

Somehow this seems close to Kleene's {\em first recursion theorem}.
Can a precise connection be made? A  gap to be closed  is that the first recursion theorem concerns computable functionals (second-order), rather than first-order functions.

\ee

\subsection{Acknowledgements} 

Thanks to NII (Tokyo), IMDEA Software (Madrid), UTS (Sydney),
and  the COLA project (University of Copenhagen) for  good environments to do this work. 
A 2013 visit to  LORIA (Nancy, France) to work with Jean-Yves Marion and Guillaume Bonfante was a great help, both in focusing this paper, and in thinking about future exciting directions, e.g., self-interpretation with only additive overhead. Thanks  for comments on form and content
to  anonymous referees and to Geoff Hamilton, Barry Jay, Torben Mogensen, Lawrence Moss, Jean-Yves Moyen and Jakob Grue Simonsen.
\vair

\fl Finally and on a broader plane, I would like {\em  to thank Dave Schmidt} for numerous wide-ranging and deep discussions of programmming languages and their semantics and implementation. These have occurred  over many years since our first contact in Kansas in the 1970s and subsequent years in Aarhus and Copenhagen.

\bibliographystyle{eptcs}

\bibliography{bib}

\nocite{*}

\end{document}